# Advances on Image Interpolation Based on Ant Colony Algorithm

Olivier Rukundo
Department of Communications and Information Sciences, Tilburg University,
Warandelaan 2,5037AB Tilburg, Netherlands
orukundo@gmail.com

Hanqiang Cao
Department of Electronic and Information Engineering, Huazhong University of Science and Technology
1037 Luoyu Road, Wuchang, 430074 Wuhan, P.R. China
caohq@hust.edu.cn

**Abstract**- This paper presents an advance on image interpolation based on ant colony algorithm (AACA) for high resolution (H.R) image scaling. The difference between the proposed algorithm and the previously proposed optimization of bilinear interpolation based on ant colony algorithm (OBACA) is that AACA uses global weighting, whereas OBACA uses local weighting scheme. The strength of the proposed global weighting of AACA algorithm depends on employing solely the pheromone matrix information present on any group of four adjacent pixels to decide which case deserves a maximum global weight value or not. Experimental results are further provided to show the higher performance of the proposed AACA algorithm with reference to the algorithms mentioned in this paper.

**Keywords**- global weight, local weight, ant colony optimization, high resolution, image interpolation

## 1. Introduction

Ant colony (optimization) algorithm, as most of the bio-inspired computational techniques, provides a very good solution to difficult optimization problems [1]. Techniques based on this nature-inspired approach have been widely applied in many engineering areas including image processing though their applications for image interpolation purposes are still few. The most common issue in image interpolation is the low-pass filtering process which reduces, to some degree, the resolution of interpolated image. However, high resolution image interpolation process has been a problem of prime importance in many fields due to its wide application in satellite imagery, biomedical imaging, particularly in military and consumer electronics domains.

In our previous work [2], ant colony optimization has been used to reinforce locally the traditional bilinear weighting scheme, in order to achieve a higher resolution interpolation. Furthermore, in [3], it has been applied to classify each wavelet coefficient into one of the Gaussian component, positive exponential and negative exponential components, before estimating the parameters of each component, but this is beyond the scope of this paper.

The local weighting scheme used in our previous work improved the peak signal to noise ratio (PSNR). However, that scheme could not avoid the influence of isotropic weighting of the conventional bilinear. Such isotropic weighting is responsible for many blurring artefacts which make the interpolated image look smoother than the original, thus reducing its resolution. As a solution, we propose a global weighting scheme which employs the pheromone matrix information solely, present on any group of four adjacent pixels, to decide which case deserves a maximum global weight value or not.

This paper is organized as follows: Part 1 the introduction; Part 2 introduces the ant colony (optimization) algorithm; Part 3 presents the proposed global weighting scheme; Part 4 gives the experimental results and Part 5 the conclusion and recommendations.

## 2. Ant colony (optimization) algorithm

Dorigo and Gambardella introduced four modifications in [4] ant system (AS) to increase its performance [5], and be able to find a very good solution to difficult optimization problems. Such modifications are: (a) a different transition rule; (b) local and global pheromone trail updates; (c) the use of local updates of pheromone trails to favor exploration; and (d) a candidate list to restrict the choice of the next node to visit.

### 2.1 Ant colony optimization state transition rule

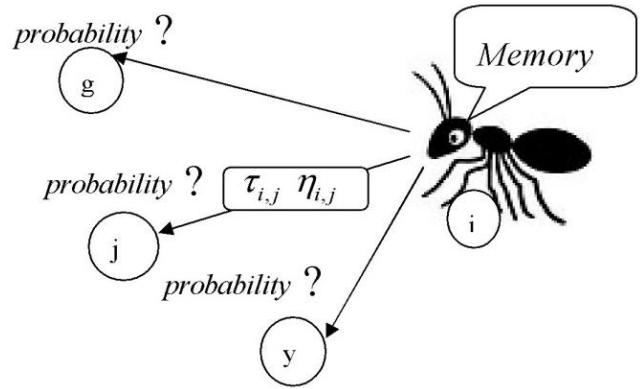

Fig.1: The selection of the node

Fig.1 shows an ant arriving in node $i$, while Table 1 shows the pseudocode of the implementation process. As shown in Fig.1, the ant chooses the next node to move to as a function of the pheromone values $\tau_{i,j}$ and the heuristic values $\eta_{i,j}$ on the arcs connecting city $i$ to the nodes $j$ the ant has not visited yet.

Table 1: A pseudocode of implementation

```
Do initialization procedures
for each iteration n = 1:N do
    for each construction_step l = 1:L do
        for each ant k = 1:K do
            Select and go to next pixel according to exploitation and
            exploration mechanism
            Apply the local pixel's updating rule
        end
    end
    Apply the global updating rule using the best visited pixels'
    pheromones
end
Decide a global weight based the best visited pixel's
pheromone information
End
```

The next node/city is chosen among the non-visited nodes (or cities) according to the pseudorandom proportional rule, in which the transition probability depends on a random variable $q$ that is uniformly distributed over [0, 1] and a parameter $q_0$.

$$j = \begin{cases} \arg\max_{f \in \Omega_i^k} \{\tau_{if}(\eta_{if})^\beta\}, & if \quad q \leq q_0 \, (Exploitation) \\ J & else \quad (Exploration) \end{cases}$$

If $q \leq q_0$, then the transition that maximizes $\tau_{i,j}(\eta)^\beta$ is chosen otherwise, and the probabilistic decision rule, Eq.(1) with $\alpha = 1$, is used. The value of $q_0$ determines the degree of exploration of the ants: with probability $q_0$, the ant chooses the transition with the highest $\tau_{i,j}(\eta)^\beta$, while with probability $1 - q_0$, it performs a biased exploration of the edges. The balance between biased exploration and pheromone exploitation can be tweaked by adjusting the value of $q_0$.

$$p_{i,j}^k = \frac{(\tau_{i,j}^\alpha)(\eta_{i,j}^\beta)}{\sum_{f \in \Omega_i^k}(\tau_{i,f}^\alpha)(\eta_{i,f}^\beta)} \quad if \quad \begin{matrix} j \in \Omega_i^k \\ \alpha = 1 \end{matrix} \tag{1}$$

In Eq.(1), $\tau_{i,j}$ is the amount of pheromone deposited for transition from state $i$ to state $j$. $\alpha \geq 0$ and $\beta \leq 1$ are parameters to control the influence of $\tau_{i,j}$ and $\eta_{i,j}$, respectively. $\eta_{i,j}$ is the desirability of the state transition $i, j$ and is equal to $1/d_{i,j}$ where $d$ is the distance in Traveling Salesman Problem (TSP). The $\Omega_i^k$ represents an acceptable neighborhood for an ant $k$ while being at city $i$ (here, the probability of selecting a city outside acceptable neighborhood is zero).

## 2.2 Ant colony optimization pheromone trail updating
### 2.2.1 Local updating
Ants use a local pheromone update rule after leaving the arc $(i, j)$ during the construction process as follows:

$$\tau_{i,j} \leftarrow (1-\varphi)\tau_{i,j} + \varphi\Delta\tau_0 \tag{2}$$

where $\varphi$, $(0 < \varphi < 1)$, $\tau_{i,j}$ is the current value of the pheromone trails at $i, j$, and $\tau_0$ is the initial value for the pheromone trails.

### 2.2.2 Global updating
The best ant reinforces its tour by depositing additional pheromone trails along the tour length

$$\tau_{i,j} \leftarrow (1-\rho)\tau_{i,j} + \rho\Delta\tau_{i,j}^{bs}, \quad \forall (i,j) \in T^{bs} \tag{3}$$

where $\Delta \tau_{i,j}^{bs} = 1/C^{bs}$ (here, $C^{bs}$ is the best-so-far tour) and $\rho$ represent the pheromone evaporation. For this case, the computational complexity is reduced from $\partial(n^2)$ to $\partial(n)$ (where $n$ is the instance being solved) by only performing the pheromone updates to the arc of $T^{bs}$ (not to all arcs as in ant system).

### 2.3 Ant colony optimization candidate list
A candidate list is a list of preferred cities to visit. Instead of examining all the cities, non-visited cities are examined first. This list is ordered by increasing distance and helps to select the next city when other cities have been visited.

### 2.4 Previous works on ant colony optimization for image interpolation
In [3], Tian, Ma and Yu proposed a wavelet-based technique, employing ant colony optimization, for image interpolation. Firstly, a J-level wavelet decomposition was applied on the input low-resolution image. The proposed TCEM model was used to formulate the statistical distribution of the above J-level wavelet coefficients and the model parameters of the proposed TCEM model were estimated using the ant colony optimization. The distribution of the desired high-pass filtered wavelet coefficient in the 0th level subband was estimated and then generated the 0th level wavelet coefficients. Finally, the (J + 1)-level inverse wavelet transform was applied to produce the final reconstructed image.

Experiments, using grayscale test images, were conducted to compare the performance of the proposed approach with that of algorithm mentioned. However, the basic wavelet based technique idea used in [3] is beyond the scope of this paper.

In [2], Rukundo, Huang and Cao applied ant colony optimization pheromone information to supplement the bilinear isotropic weighting scheme in order to achieve improved (or higher) resolution results. Here, ant colony optimization was used to construct the pheromone matrix and then find the pixels on which more pheromone information was deposited. This pheromone matrix information was used to strengthen the traditional bilinear algorithm weighting scheme shown in Eq.(4).

$$P_{x,y} = w_4 P_{u,v} + w_2 P_{u,v+1} + w_3 P_{u+1,v} + w_1 P_{u+1,v+1} \tag{4}$$

where $(x, y)$ represents the coordinates of unknown-value location, $(u, v)$, $(u+1, v)$, $(u, v+1)$, and $(u+1, v+1)$ are coordinates of each of the four (known-value) pixels surrounding $(x, y)$ location.

$$\begin{aligned} w_1 &= \{[x-u][y-v]\}, w_2 = \{[x-(u+1)][y-(v)]\}, \\ w_3 &= \{[x-(u)][y-(v+1)]\}, w_4 = \{[x-(u+1)][y-(v+1)]\} \end{aligned} \tag{5}$$

Our previous OBACA algorithm used the pheromone matrix information (obtained after both updates were performed), to create additional but necessary weights. In other words, the pheromone matrix information was assigned on every pixel with reference to the neighboring pixels intensity variations. The pheromone information attached to each pixel became its weight, to locally supplement the traditional bilinear weighting scheme, as shown in Eq.(6),

$$P_{x,y} = p_4 w_4 P_{u,v} + p_2 w_2 P_{u,v+1} + p_3 w_3 P_{u+1,v} + p_1 w_1 P_{u+1,v+1} \tag{6}$$

where $p_4, p_2, p_3, p_1$ represent the pheromone information at $(u,v)$, $(u+1,v)$, $(u,v+1)$ and $(u+1,v+1)$, respectively. In this way, OBACA achieved higher PSNR or image quality when compared to that of the traditional bilinear. However, it could not avoid the influence of isotropic weighting of the conventional bilinear which is responsible for smoothing/blurring the output/interpolated images, thus reducing their resolutions. This is the aim of our research to develop a global weighting scheme whose basic function is shown by Eq.(7), where $w_g$ the global weight produced by ant colony optimization algorithm.

$$P_{x,y} = w_g (w_4 P_{u,v} + w_2 P_{u,v+1} + w_3 P_{u+1,v} + w_1 P_{u+1,v+1}) \qquad (7)$$

Furthermore, the developed weighting scheme considers "the characteristics of pheromone information locality" in order to assign that global weight thus achieves a higher resolution image interpolation.

### 3. The proposed algorithm

As shown by Eq. (7), $w_g$ is the global weight which is applied to a group of pixels surrounding directly an unknown-value location. This global weight is a function of pheromone matrix information on that/each group of four pixels. Now, the problem is how to find the pheromone information present on each group. To solve this problem, we started by constructing the pheromone matrix and then we checked possibilities of the partial or full simultaneous presence of the pheromone trails information at the four pixels/locations surrounding directly the unknown-value location $P_{x,y}$.

### 3.1 Pheromone matrix construction

Initially a pheromone value $\tau_{init}$ is assigned to every pixel location (see Fig.3-(a)) and, ants are randomly distributed on image before moving artificial ants for a certain number of steps on that image.

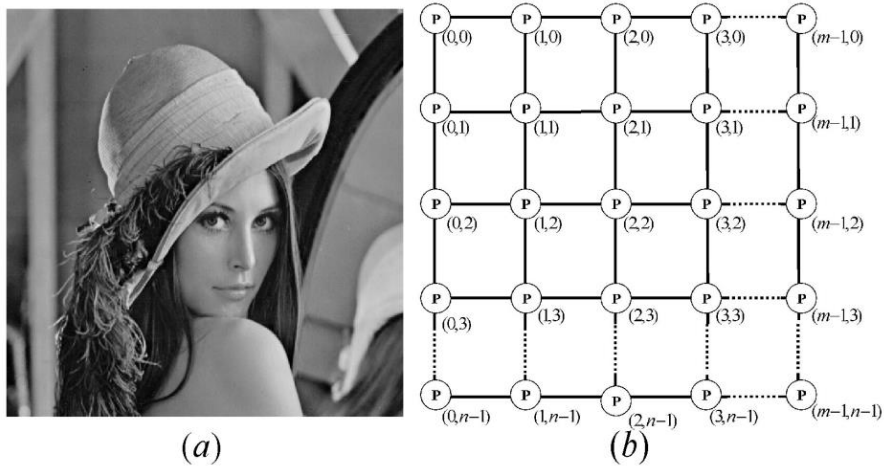

(a) (b)

Fig.2: Fig.2-(a) shows Lenna image. Fig.2-(b) is a representation in a two-dimensional matrix in which each image pixel location is represented by the pheromone information $P$.

The ants' movements are/will be steered by the intensity variation of the grayscale pixels (within a permissible range of 8-connectivity neighborhood around $P_{(i,j)}$, as shown in Fig.3-(b).

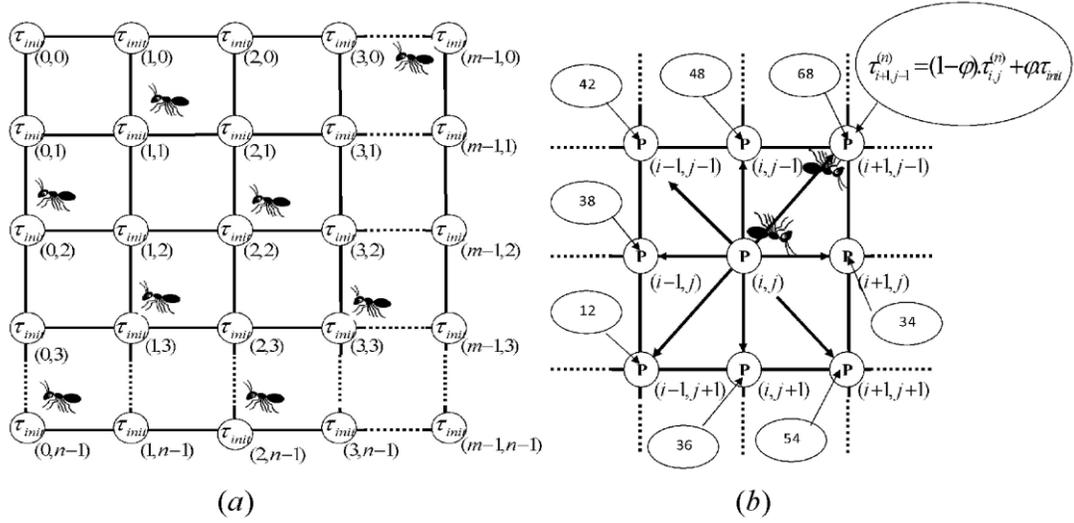

Fig.3: (a) shows the initial pheromone matrix look. Initially, the pheromone value for each matrix element is set to a non-zero constant; (b) shows the 8-connectivity neighborhood and possible ant's direction movement.

The $n^{th}$ ant's move from a departing pixel's position $(i_o, j_o)$ located within acceptable (8-connectivity) neighborhood depends on the attractiveness $\eta_{i,j}$ of any adjacent pixel [e.g. $i, j$ ] and the pheromone trail information $\tau_{i,j}^{(n-1)}$ on it. Therefore, an ant will make a $n^{th}$ move from any pixel $(i_o, j_o)$ to an adjacent pixel $(i, j)$ according to the pseudorandom proportional rule. The transition probability will be given by Eq.(8)

$$p_{(i_o, j_o),(i,j)}^{(n)} = \frac{(\tau_{i,j}^{(n-1)})^\alpha (\eta_{i,j}^{(n-1)})^\beta}{\sum_{(i,j)\in \Omega(i_o, j_o)} (\tau_{i,j}^{(n-1)})^\alpha (\eta_{i,j}^{(n-1)})^\beta} \qquad (8)$$

where $\tau_{i,j}^{(n-1)}$ is the pheromone value on the pixel $(i, j)$; $\Omega_{(i_o, j_o)}$ is an acceptable neighborhood for the pixel $(i_o, j_o)$; $\eta_{i,j}$ is the heuristic information on the pixel $(i, j)$; the pheromone and heuristic information constants $\alpha$ and $\beta$ are always positive. Each time an ant visits a pixel, it automatically performs a local update on the adjacent pixel. The amount of pheromone on the pixel $(i, j)$ at the $n^{th}$ iteration, $\tau_{i,j}^{(n)}$ is calculated using the local pheromone update Eq. (9)

$$\tau_{i,j}^{(n)} = (1-\varphi).\tau_{i,j}^{(n)} + \varphi.\tau_{init} \qquad (9)$$

where $\varphi \in [0,1]$ is the pheromone decay coefficient. In each iteration, the pheromone trail value changes because local updates are provided together with the solution construction process. As mentioned, the ant's permissible range of movements is situated within the 8-connectivity neighborhood, see Fig.3 (b). An ant can move to an adjacent pixel on the condition that it moves to a pixel that was not recently visited by any other ants. This condition is backed by an artificial memory assigned to each ant so that it can keep records of every visited pixel. After all ants have completed the construction process, a global pheromone update is performed only on the pixels that have been visited by at least one ant according to Eq.(10)

$$\tau_{i,j}^{(n)} = (1-\rho).\tau_{i,j}^{(n-1)} + \rho.\sum_{K=1}^{K}\Delta\tau_{i,j}^{(k)} \qquad (10)$$

where $\Delta\tau_{i,j}^{(k)}$ is the amount of pheromone deposited by the $k^{th}$ ant on the pixel $(i,j)$. This amount $\Delta\tau_{i,j}^{(k)}$ is equal to the average of heuristic information associated with the pixels that belong to the tour of the $k^{th}$ ant if the pixel $(i,j)$ was visited by the $k^{th}$ ant in its current tour, otherwise the amount $\Delta\tau_{i,j}^{(k)} = 0$; $\rho$ is the pheromone evaporation rate; $K$ is the number of ants which can be determined by Eq.(11).

$$K = \sqrt{input\ image\ width * input\ image\ height} \qquad (11)$$

As the value for $\Delta\tau_{i,j}^{(k)}$ depends on the heuristic information, $\eta_{i,j}$ at the $(i,j)$ can be given by Eq. (12)

$$\eta_{(i,j)} = \frac{V_c(P_{i,j})}{V_{max}} \qquad (12)$$

where $P_{(i,j)}$ is the intensity value of a pixel at $(i,j)$, $V_{max}$ is the normalization factor; $V_c(P_{i,j})$ is the function that operates on the 8-connectivity pixels 'neighborhood. The value of that function depends on the variation of the pixel's intensity values and can be given by Eq.(13)

$$V_c(P_{i,j}) = |P_{(i-1,j-1)} - P_{(i+1,j+1)}| + |P_{(i+1,j-1)} - P_{(i-1,j+1)}| \\ + |P_{(i,j-1)} - P_{(i,j+1)}| + |P_{(i-1,j)} - P_{(i+1,j)}| \qquad (13)$$

### 3.2 Max and mean global weighting scheme

This is a very important step of our proposed algorithm. The max function returns the greatest element value in a data set whereas the mean returns the average of elements values. Eq.(10) gives the pheromone value or level on mostly visited pixel locations. However, it has been experimentally upgraded to Eq.(14) in order to yield viewable results.

$$p_{(i,j)} = \exp(\tau_{i,j}^n) \qquad (14)$$

Intuitively, assigning a direct the global weight, as shown by Eq.(7), would not permit to reduce the low-pass filtering process which reduces image resolution. However, in order to achieve higher resolution or quality image interpolation, we have introduced a weighting scheme which employs solely the Eq.(14) pheromone information, present on any group of four adjacent pixels, to decide which interpolated pixel that needs the maximum global weight or not. Now, with reference to Fig.4 and Eq.(12), we can estimate the desirability of an ant to select the next pixel using the function $V_c(P_{i,j})$.

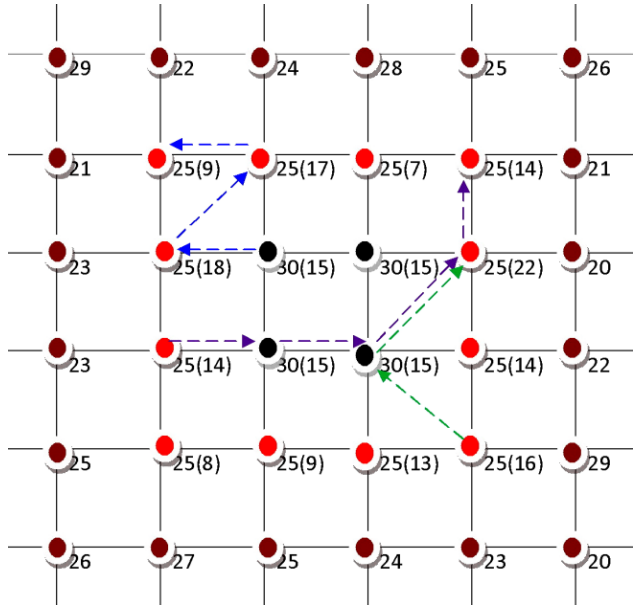

Fig.4: The estimation of an ant to select the next pixel

For instance, Fig.4 shows that an ant belonging to the upper left black pixel dot 30(15), where 30 is the pixel value and 15 is $V_c(P_{i,j})$ value, has eight choices in which the best one (i.e. one having the highest $V_c(P_{i,j})$ value) is acceptable, if and only if it has not been chosen by any (other) ants. In this way, the ant will follow the blue, purple or green arrows depending on where it was initially put or randomly distributed. Their moves along those arrows shows that the pheromone information can be present (at least) at three, two and one locations in a group (of four pixels), because every time an ant moves, it drops the pheromone trails on the visited pixel or location.

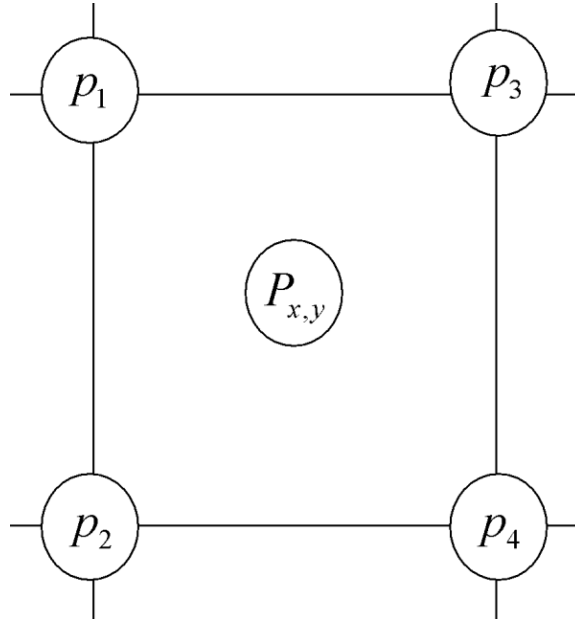

Fig.5: Pheromones trails around the unknown-value location $P_{x,y}$

Now, the next problem is how reasonably to assign the global weight in each (simultaneous pheromone trails presence) case. To settle this, we label $p_1$, $p_2$, $p_3$ and $p_4$ as the pheromone information on each pixel position surrounding directly the unknown-value location $P_{x,y}$ shown in Fig.5, and that $A$, $B$, $C$ and $D$ are different levels or value that $p_1$, $p_2$, $p_3$ and $p_4$ can have. Let us now make $S$ a set containing the pheromone trails information $p_1$, $p_2$, $p_3$ and $p_4$ as shown by Eq.(15).

$$S = [p1, p2, p3, p4] \tag{15}$$

In order to reduce reasonably the low-pass filtering process in our algorithm we used max and mean functions to assign reasonably the global weight with reference to possible partial or full simultaneous presence of the pheromone trails information at the four locations surrounding directly the unknown-value location $P_{x,y}$.

- **Assumption 1**

When $p_1 = p_2 = p_3 = p_4 = A$ (all pheromone trails information around $P_{x,y}$ have the same level).

$$Max(S_1) = A \tag{16}$$

And

$$Mean(S_1) = A \tag{17}$$

Here, there is no special requirement for assigning the highest level of pheromone trails. Therefore, any of two weighting procedures can be used, my experimental choice is given by Eq.(18)

$$w_g = Mean(S_1) \tag{18}$$

- **Assumption 2**

When $p_1 = p_2 = p_3 = A$ and $p_4 = B$ (one of the four pheromone trails information around $P_{x,y}$ has a different level).

$$\begin{cases} Max(S_2) = A & \text{if } p3 > p4 \\ Max(S_2) = B & \text{otherwise} \end{cases} \tag{19}$$

And

$$Mean(S_2) \neq A \neq B \tag{20}$$

Here, a highest level of pheromone trails can be assigned reasonably only when $p_3 > p_4$, since it is only in this case where you can find one location having the lowest pheromone trail level. For this reason, the global weight is given by Eq.(21)

$$w_g = Max(S_2) \tag{21}$$

- **Assumption 3**

When $p_1 = p_2 = A$, $p_3 = C$ and $p_4 = B$ (only two of the four pheromone trails information around $P_{x,y}$ have the same level).

$$\begin{cases} Max(S_3)=A & if \quad p2>\{p3,p4\} \\ Max(S_3)=C & if \quad p2<p3 \\ Max(S_3)=B & if \quad p2<p4 \end{cases} \qquad (22)$$

And

$$Mean(S_3) \neq A \neq B \neq C \qquad (23)$$

A highest level of pheromone trails can be assigned reasonably only when $p_2 > p_3$ & $p_2 > p_4$, because it is in this case where you can only find a biggest number of locations with the highest pheromone trail level, at the same time. Therefore, the max weighting is given by Eq.(24)

$$w_g = Max(S_3) \qquad (24)$$

- **Assumption 4**

When $p_1 = A$, $p_2 = D$, $p_3 = C$ and $p_4 = B$ (all pheromone trails information around $P_{x,y}$ have different levels).

$$\begin{cases} Max(S_4)=A & if \quad p1>\{p2,p3,p4\} \\ Max(S_4)=D & if \quad p2>\{p1,p3,p4\} \\ Max(S_4)=C & if \quad p3>\{p2,p1,p4\} \\ Max(S_4)=B & if \quad p4>\{p2,p3,p1\} \end{cases} \qquad (25)$$

And

$$Mean(S_4) \neq A \neq B \neq C \neq D \qquad (26)$$

Here, a mean-based weighting procedure is preferable in order to avoid any possible influence of a strong noise intrusion. Therefore, the global weight is given by Eq.(27)

$$w_g = Mean(S_4) \qquad (27)$$

- **Assumption 5**

When $p_1 = p_2 = A$, $p_3 = p_4 = B$ (every two pheromone trails information around $P_{x,y}$ have the same level).

$$\begin{cases} Max(S_5)=A & if \quad p2>p3 \\ Max(S_5)=B & if \quad p2<p3 \end{cases} \qquad (28)$$

And

$$Mean(S_3) \neq A \neq B \qquad (29)$$

A highest level of pheromone trails can be assigned reasonably only when $p_2 > p_3$ & $p_2 < p_3$ and therefore, the max weighting is given by Eq.(29)

$$w_g = Max(S_5) \qquad (30)$$

## 3.4 Parameter selection

The ant colony optimization has many parameters that need to be properly selected in order to increase the strength of the proposed algorithm. In this regard, the proper selection of parameters $\alpha$ and $\beta$ helps to determine the relative influence of the pheromone trail and the path visibility. Information in image content is always more important than in pheromone trail, thus, $\beta > \alpha$ is a general selection and our experiments adopt $\alpha = 1$, $\beta = 2$ as in [3]. Other parameters such as $\tau_{init}$, $\varphi$ and $\rho$ determine the pheromone trail change, where $\tau_{init}$ is the initial value for the pheromone trails and this value is always non-zero though very close to zero. $\varphi$ and $\rho$ are adaptation parameters. Our experiment adopt $\tau_{init} = 0.0001$, $\varphi = 0.00001$ and $\rho = 0.1$. Finally, the number of iterations, ants and memory size are also important parameters which deserve a proper selection. Note that a large number of iterations tend to increase the image edges sharpness but at very high computational load plus a possibility of generating too many false edge pixels. Furthermore, with a higher number of iterations, the change in image quality is not significant as it can be imagined intuitively. Artificial ant's memory size can be determined by Eq.(31) to ensure that each ant is assigned a big enough memory to record all '*visitable*' pixels.

$$Memory\ size = \frac{m*n}{K} \qquad (31)$$

Experimentally, we found that the number of iterations ranging from 1 to 10, the number of ants given by Eq.(11) and the memory size given by Eq.(31) are adequate parameter values for yielding interpolated images in their most pleasant way within a tolerable computational load.

## 3.3 Summary of the proposed algorithm

Fig.6 shows a summary of the proposed AACA algorithm. The proposed algorithm includes two important steps namely, the pheromone matrix construction and global weighting scheme.

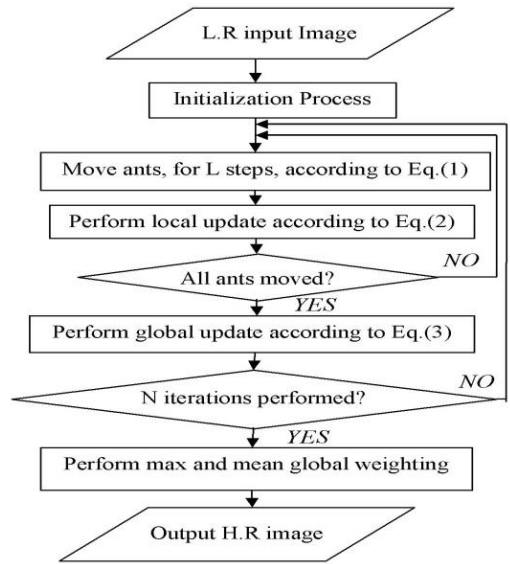

Fig.6: Summary of the proposed algorithm

The proposed approach provides better results than our previously proposed methods as well as the interpolation algorithms mentioned in this paper. The novelty in this approach is a weighting scheme involved in its process. This scheme is based on global weight assignment rather than local. It uses the max and mean functions to adapt adequately the global weight to an interpolated pixel with reference to partial or full simultaneous presence of the pheromone trails information (at the four locations). This is done to reduce reasonably the low-pass filtering process thus achieve a H.R image interpolation.

## 4. Experimental results

This section presents the results from the experiments conducted. The performance of the proposed approach was compared with that of the previous works, two traditional interpolation algorithms executed by MATLAB toolbox and new edge directed interpolation algorithm proposed/whose details are given in [6]. Two image quality assessment measures used are: Peak Signal to Noise Ratio (PSNR) and Mean Squared Error (MSE). A higher PSNR would normally indicate the higher quality of the output image whereas a higher MSE would indicate the interpolator's weakness to reconstruct faithfully an image. Table 2 and Table 3 provide the MSE and PSNR values for interpolated *Lenna*, *Cameraman*, *Lake* and *Peppers* test images. However, Fig.7, Fig.8, Fig.9 and Fig.10 show these images after being interpolated using the bilinear, bicubic, OBACA, Nearest Neighbor Value Interpolation (NNV) proposed in [7] and AACA interpolation algorithms presented in this paper. Furthermore, Fig.11, Fig.12, Fig.13 and Fig.14 show the portions of *Lenna*, *Lake*, *Peppers* and *Cameraman* images interpolated using New Edge Directed Interpolation (NEDI) and AACA. The PSNR and MSE were provided in the caption of Fig.11, Fig.12, Fig.13 and Fig.14.

Table 2: MSE of the interpolated images (128→512)

| Images | Methods | Values |
|---|---|---|
| Cameraman | Bilinear | 25.2190 |
| | Bicubic | 24.9344 |
| | OBACA | 21.9463 |
| | NNV | 20.9192 |
| | AACA | 17.4473 |
| Lenna | Bilinear | 23.0509 |
| | Bicubic | 22.1158 |
| | OBACA | 15.9800 |
| | NNV | 14.9963 |
| | AACA | 10.6305 |
| Lake | Bilinear | 43.2509 |
| | Bicubic | 42.5583 |
| | OBACA | 39.8727 |
| | NNV | 38.7641 |
| | AACA | 33.6966 |
| Peppers | Bilinear | 27.1935 |
| | Bicubic | 26.8901 |
| | OBACA | 22.9027 |
| | NNV | 21.2183 |
| | AACA | 17.7068 |

Table 3: PSNR of the interpolated images (128→512)

| Images | Methods | Values (dB) |
|---|---|---|
| Cameraman | Bilinear | 34.1135 |
| | Bicubic | 34.1628 |
| | OBACA | 34.7172 |
| | NNV | 35.5267 |
| | AACA | 35.7135 |
| Lenna | Bilinear | 34.5039 |
| | Bicubic | 34.6838 |
| | OBACA | 36.0950 |
| | NNV | 36.1891 |
| | AACA | 37.8653 |
| Lake | Bilinear | 31.7708 |
| | Bicubic | 31.8410 |
| | OBACA | 32.1240 |
| | NNV | 32.1377 |
| | AACA | 32.8549 |
| Peppers | Bilinear | 33.7862 |
| | Bicubic | 33.8349 |
| | OBACA | 34.5319 |
| | NNV | 34.5559 |
| | AACA | 35.6494 |

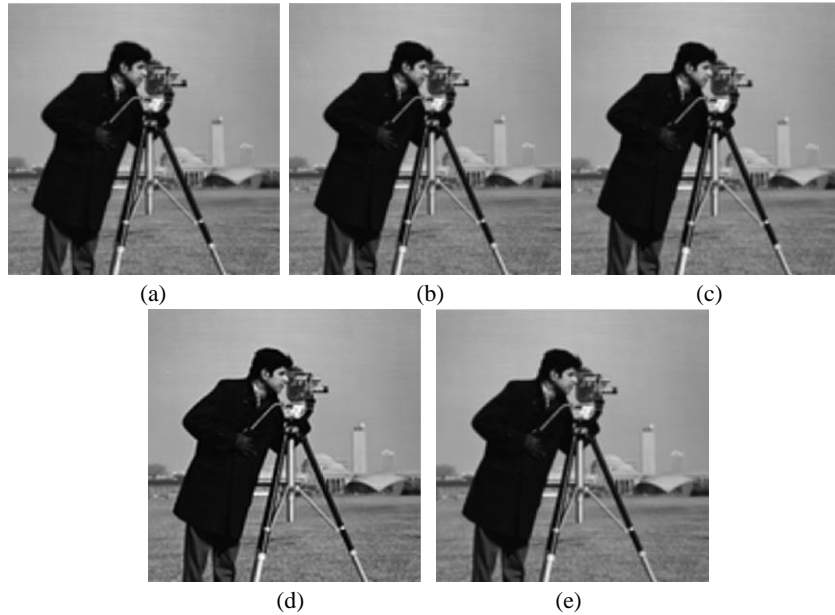

(a)　　　　　(b)　　　　　(c)

(d)　　　　　(e)

Fig.7: (a) interpolated image by AACA; (b) interpolated image by NNV; (c) interpolated image by OBACA; (d) interpolated image by bicubic; and (e) interpolated image by bilinear algorithm

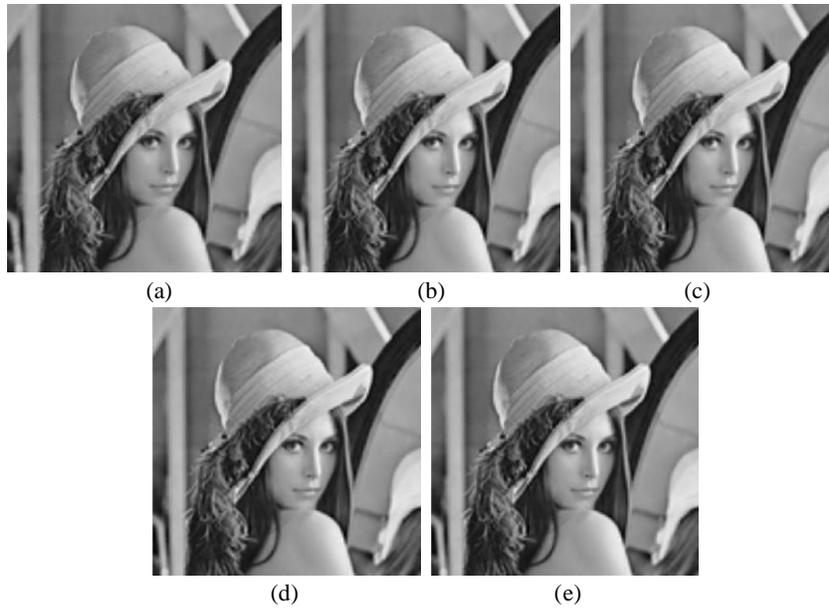

Fig.8: (a) interpolated image by AACA; (b) interpolated image by NNV; (c) interpolated image by OBACA; (d) interpolated image by bicubic; and (e) interpolated image by bilinear algorithm

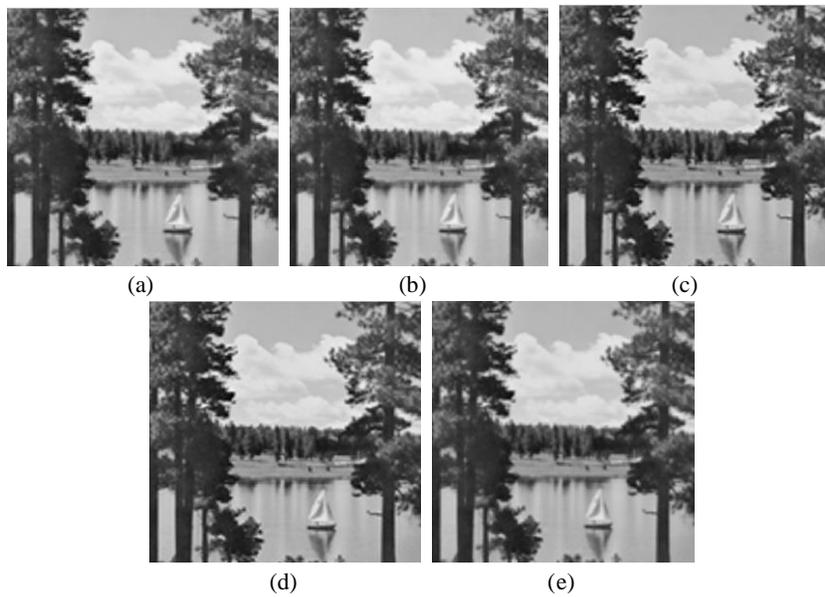

Fig.9: (a) interpolated image by AACA; (b) interpolated image by NNV; (c) interpolated image by OBACA; (d) interpolated image by bicubic; and (e) interpolated image by bilinear algorithm

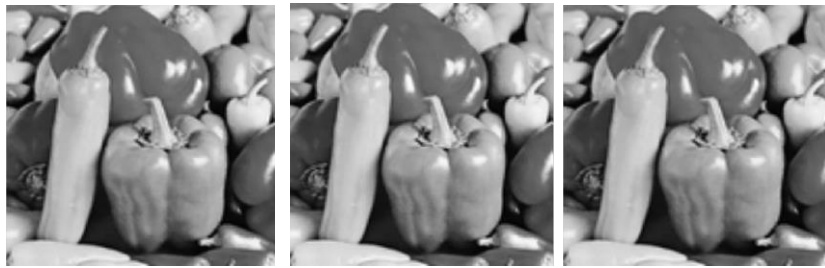

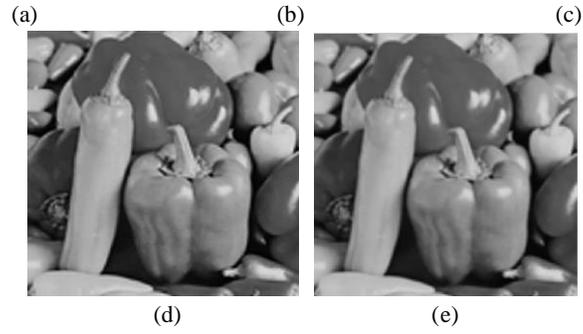

(a)          (b)          (c)

(d)          (e)

Fig.10: (a) interpolated image by AACA; (b) interpolated image by NNV; (c) interpolated image by OBACA; (d) interpolated image by bicubic; and (e) interpolated image by bilinear algorithm

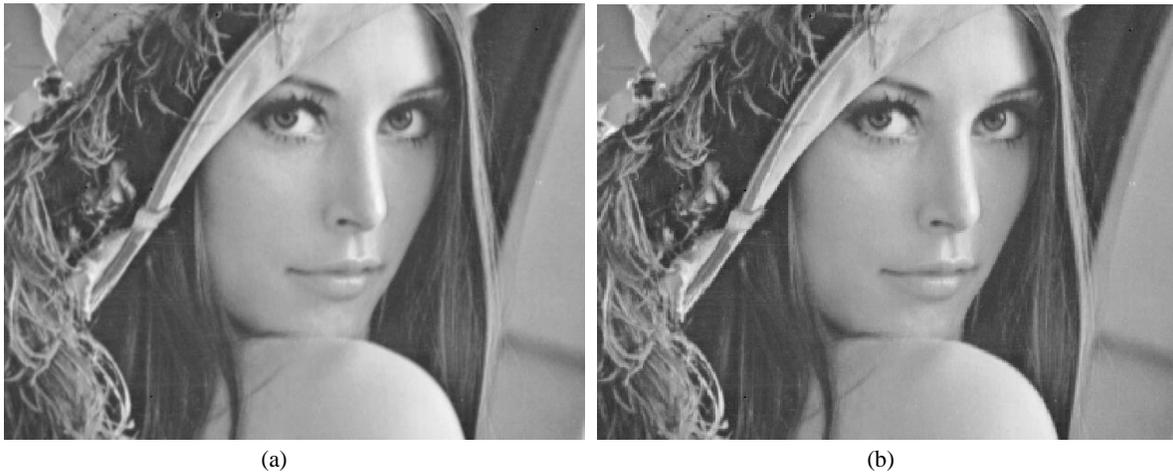

(a)          (b)

Fig.11: (a) interpolated image by AACA (PSNR=39.2180, MSE=7.7853); (b) interpolated image by NEDI (PSNR=37.1192, MSE=12.6229)

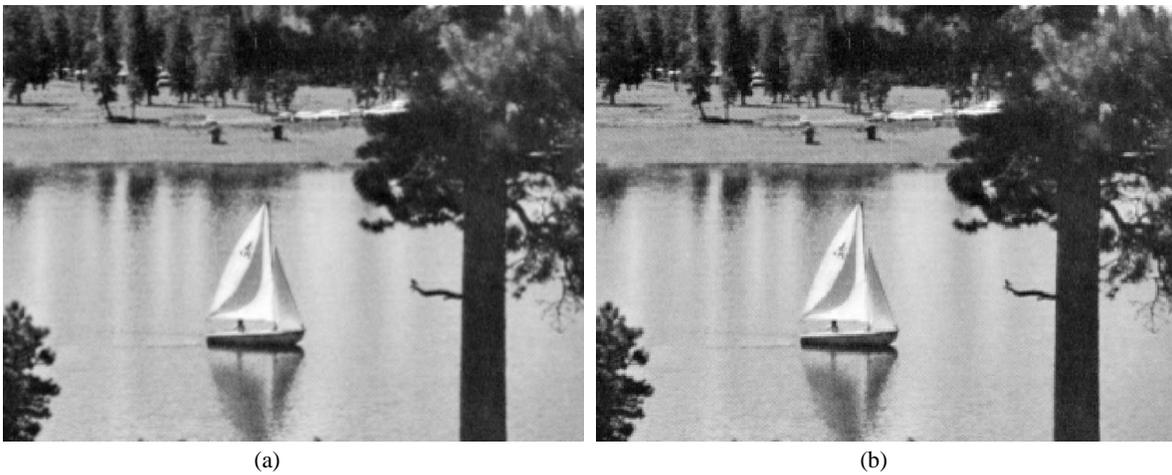

(a)          (b)

Fig.12: (a) interpolated image by AACA (PSNR=37.9615, MSE=10.3976); (b) interpolated image by NEDI (PSNR=35.9885, MSE=16.3769)

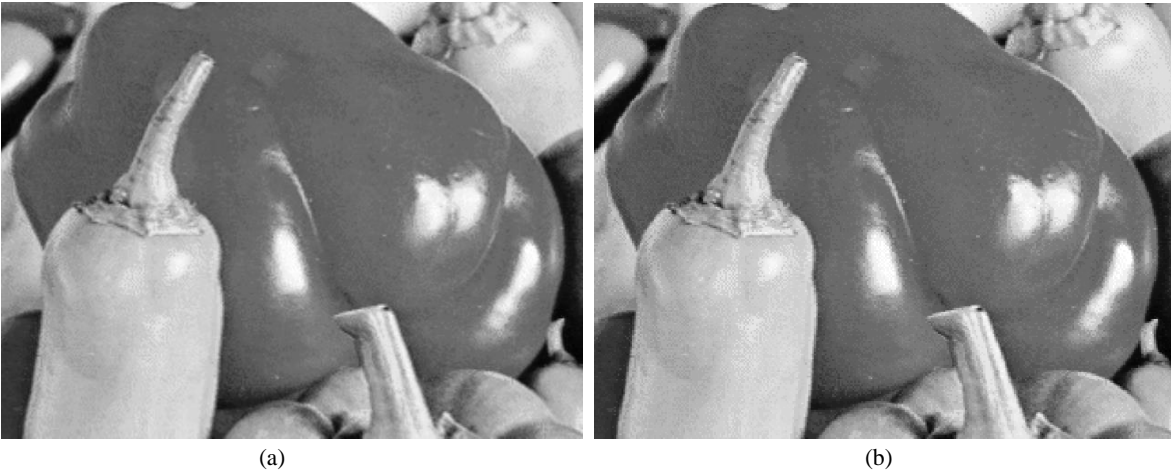

(a)                          (b)

Fig.13: (a) interpolated image by AACA (PSNR=40.1928, MSE=6.2201); (b) interpolated image by NEDI (PSNR=37.8635, MSE=10.6348)

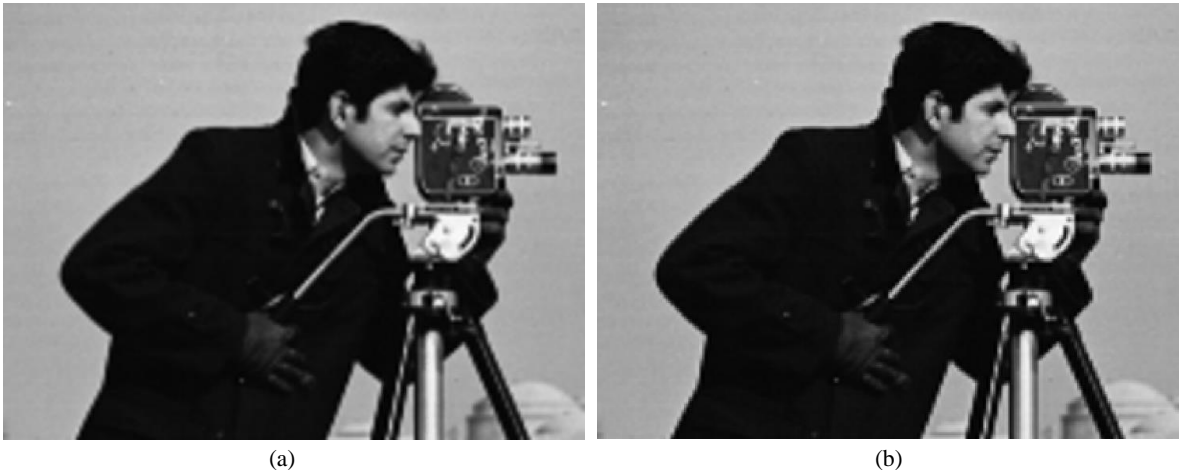

(a)                          (b)

Fig.14: (a) interpolated image by AACA (PSNR=40.1144, MSE=6.3334); (b) interpolated image by NEDI (PSNR=37.4861, MSE=11.6003)

## 5. Conclusions and recommendations

An advance on image interpolation based on ant colony algorithm has been presented in this paper. Unlike our previous OBACA algorithm, which employed a local weighting scheme, the presented AACA algorithm used however a novel global weighting scheme. The strength of the proposed global weighting depended on employing solely the pheromone matrix information, present on any group of four adjacent pixels, to decide which case deserves a maximum global weight value or not. The use of max and mean global weight values has shown that the proposed AACA algorithm is able to reduce the interpolation errors with reference to the original image. This was proved by the experiments conducted on full and partial *Cameraman*, *Lenna*, *Lake* and *Peppers* test images. More particularly, it has been shown by higher quality/resolution images yielded by AACA algorithm than other algorithms mentioned in this paper. The future development of the proposed approach may be devoted to settling the computational load issue by restricting the movements of ants on some parts of an image.

**Authors' contributions**

O. Rukundo made substantial contributions to conception and design of the advanced strategy of the algorithm proposed. H.Q. Cao revised the manuscript critically to meet the expected standards in scientific publishing. Both authors read and approved the final manuscript.

**Competing interests**

Authors declare that they have no competing interests.

**Acknowledgement**

This work was supported by the National Anti-counterfeit Engineering Research Center and National Natural Science Foundation of China (N0: 60772091). Author would like to thank the reviewers and editor for their helpful comments.